\documentclass[twocolumn,showpacs,nobibnotes,nofootinbib,amsmath,amssymb]{revtex4}
\usepackage[cp1251]{inputenc}
\usepackage[T2A]{fontenc}
\usepackage[russian,english]{babel}
\usepackage{graphicx}
\usepackage{dcolumn}
\usepackage{bm}
\usepackage{color}
\begin{document}

\title{A RELATIVE INFORMATION APPROACH TO FINANCIAL TIME SERIES ANALYSIS
USING BINARY $N$-GRAMS DICTIONARIES}

\author{Igor Borovikov}
\affiliation{Nekkar.net: Int.\,Labs, 51 Common Ln., Foster City CA 94404, USA} \email{igor.borovikov@gmail.com}

\author{Michael G.\,Sadovsky}
\affiliation{Institute of computational modelling of SB RAS;\\ 660036 Russia, Krasnoyarsk,
Akademgorodok.} \email{msad@icm.krasn.ru}

\begin{abstract}
Here we present a novel approach to statistical analysis of financial time series. The approach is based on $n$-grams frequency dictionaries derived from the quantized market data. Such dictionaries are studied by evaluating their information capacity using relative entropy. A specific quantization of (originally con\-tinuous) financial data is considered: so called binary quantization. Possible applications of the proposed technique include market event study with the $n$-grams of higher information value. The finite length of the input data presents certain computational and theoretical challenges discussed in the paper. also, some other versions of a quantization are discussed.
\end{abstract}

\date{\today}
\maketitle

\newcounter{N}

\section{Introduction}
Currently, a world economy concentrates at the securities turn-over; this issue becomes tends to be both global, and essential. This turn-over machinery involves a huge number of people and various institutions; a lot has been said, and would be said more concerning the studies of the peculiarities of the behaviour of those entities and people. Basically, a key idea of any research targeted to financial data mining consists of two points: the former is to figure out some order in these data, and the latter is to elabourate a number of indicators able to predict some (important) events at the stock markets.

Mathematical modelling and relevant mathematical methods were widely applied in this area; this field of studies is knows as \textsl{technical analysis} and has a good history behind with a developed culture of doing research here. A publication abundance falls beyond any comprehensive analysis, thus we shall not pretend to provide it, while some classical works (see~\cite{fama69,Lo}) definitely should be mentioned.

Money flow, with neither respect to its specific form, could be considered as a time series, either discrete, or continuous, and relevant mathematical techniques of the analysis could be implemented. A good starting reading could be found in \cite{Lo,tsay}; relevant but somehow too enthusiastic introduction reading in Russian is provided by \cite{kijaniza}. Finally, the classics by J.\,Murphy \cite{murphy} must be mentioned as a key reading for anyone interested in the foundations of the subject\footnote{There are the translations into Russian of his books.}.

Basically, various techniques \cite{pettingil} and approaches of (linear) statistics analysis and probability theory \cite{1} are implemented. Empirical studies \cite{JF87} are also an important contribution to that field. Russian researchers also deal with the problem mentioned above: mathematical techniques are widely used in such studies \cite{tkachenko,kovalenja,pashch,sychev}. Advanced techniques of the data body decomposition is considered in \cite{belorus,nekrasova}. An attempt to figure out some indicative issues to forecast the market dynamics behaviour is proposed in \cite{prognoz}.

Here we follow the approach initially developed for the analysis of genetic texts in the pioneering works \cite{n1,n2}; this approach seems to be novel, for financial series analysis. The main idea is to build $n$-gram frequency dictionaries $D(n)$ from a sufficiently large input text(s) for $n$-grams of a different length $n$. The entropy-maximization procedures described in the cited works yields the new dictionaries $D^{k}(n+k)$ from $D(n)$, $k=1, 2, \ldots$\,. These are called reconstructed (or lifted) dictionaries. The Kullback-Leibler divergence between reconstructed $D^k(n)$ and the original $D(n)$ dictionaries for the same $n$ gives a \emph{relative} information capacity of the input text\footnote{Actually, the information capacity is defined for a frequency dictionary rather than for a text; further we shall not distinguish this point.} for the $n$-grams of length $n$. Such analysis also results in the detection of ``divergent'' $n$-grams responsible for ``higher information content''. The definition of information capacity introduced this way is not equivalent to the Shannon's or Boltzmann's classical ones, which are based on \emph{absolute} entropy of the text. It is worth noting that the $n$-grams frequency dictionary approach is not using any explicit assumptions regarding the text, like Markovian property, or alphabet letters distribution.

This paper discusses the application of the outlined relative information methods to the financial time series. These methods were developed for analysis of genetic texts which can be viewed as rather long words in a specific finite alphabet. The values of the time series in finance are usually represented by real numbers. To transfer the techniques developed for the genetic texts to financial time series we need to quantize real-valued series and produce words in some suitable alphabet. There are many ways of doing that and we describe some of them in the next section.

Next, following the cited works, we explain information capacity defined with the $n$-grams frequency dictionaries. This leads to selecting the optimal length $l_{\textrm{opt}}$ of $n$-grams for further analysis as the length yielding the maximum of (normalized) information content over the all $n$-gram lengths and possibly other parameters of the method. The intuition behind this is that for the optimal set of parameters, the $l_{\textrm{opt}}$-grams are the least predictable ones. We attempt to connect them to significant market events and/or trends. The normalization of the information content turns out to be more important task than in the case of applications to bioinformatics because usually financial time series result in much shorter input texts than a typical sequenced genome. 

\section{$N$-grams Dictionaries from Time Series}\label{basmodel}
To avoid an ambiguity, we shall use the term \textsl{ticker} when talking about a security like company shares, ETFs or indices (e\,.g. GOOG, YANG or $\hat{}~$DJI). We will reserve the terms \textsl{letter} and \textsl{symbols} (to be used interchangeably) for the elements of the alphabet we are going to construct.

We consider the simplest case of a financial time series, namely Adjusted Close daily price on a ticker\footnote{The source of the data used throughout this work is the publicly available financial data from \emph{Yahoo!}Finance unless indicated otherwise.} denoted by $z(t)$, from which we calculate either $\log$- or simple returns $p(t)$: \[p(t) = \log\left(\dfrac{z(t)}{z(t-1)}\right) \approx \dfrac{z(t)}{z(t-1)} - 1\,\,.\]
Here Adjusted Close price $z(t)$ is a real number and $t$ is (trading) day treated as an integer index. The choice of simple returns over log-returns is not critical for this work so we will not distinguish them further.

\subsection{Constructing texts from time series}
To apply the $n$-grams-based methods, we have to represent time series $p=p(t)$ as a (very long) text from an alphabet $\aleph$. We shall call it an \emph{input text}. The letters of the alphabet encode quantized values of $p(t)$. The choice of mapping $R \to \aleph$ (here $p(t) \in R$) and the choice of the alphabet $\aleph$ are the parameters of method. This paper is concentrated on the special case of the mapping into a binary alphabet $\aleph = \{0, 1\}$. Some other types of mappings will discussed later (see Section~\ref{discussion}). In particular, we focus on alphabets with even number of letters corresponding to different multiples of up-ticks and down-ticks in the price movement.

\newtheorem{definition}{Definition}
\begin{definition}
A finite alphabet $\aleph_N$ of the cardinality $\|\aleph\| = 2N > 0$ is called an \emph{output alphabet} if it is ordered by bijective mapping to the set of integers $Z_N = \{-N,-(N-1), \dots, -2, -1, 1, 2, \dots, N-1, N \}$ (note the absence of $0$). The mapping $X{:}\ Z_N \to \aleph_N$ is called \emph{indexing}.
\end{definition}

Binary dictionary is a special case: it can be represented with the digits $\{0, 1\}$ where $0$ corresponds to the negative values of $p$ and $1$ corresponds to the positive values. Note that quantization includes clamping of input values. In the most straightforward approach, this results in mapping of the entire ranges $(-\infty,P_{-N}]$ and $[P_m, +\infty)$ to the first and the last letters of the alphabet, correspondingly. This is useful when we need to limit the alphabet cardinality introducing no additional complicated non-linearity into quantization mapping. The binary alphabet is a trivial example of clamping with $N=1$ and $P_{-1}=P_1 = 0$.

To summarize, the base procedure that generates an input text from a series $z(t)$ of Adjusted Close prices consists of the three steps:
\begin{list}{\arabic{N})}{\topsep=0cm \itemsep=0cm \listparindent=1cm \parsep=0pt \usecounter{N}}
\item Convert prices time series $z(t)$ to ($\log$-) returns $p(t)$,
\item Specify the output alphabet $\aleph$ and the quantization mapping $Q{:}\ R \to \aleph$, 
\item Quantize $p(t)$ to obtain the text $\mathsf{T}=\left\{Q(p)\right\}$.
\end{list}

\subsection{Dictionaries from the Input Text}
Here we describe a construction of various useful frequency dictionaries of $n$-grams from the input text.

Given an input text $\mathsf{T}$ of a finite length $L$, first we build natural frequency dictionary $D(n)$ by counting all $n$-grams occurrences $C_w$ for each $n$-gram $w$ in the text $\mathsf{T}$. It yields a set of pairs $(w,C_w)$. Let $C_*$ be the total number of $n$-grams in $\mathsf{T}$. Obviously, $C_* = |\mathsf{T}| - n$, where $|\mathsf{T}|$ is the text length $L$. Normalization by $C_*$ gives the \emph{frequency} of the $n$-gram $w$: $f_w = C_w/C_*$.

\begin{definition}
The (natural) \emph{frequency dictionary} $D(n)$ of the text $\mathsf{T}$ is the set of all pairs $\{(w, f_w)\}$ where $w$ are unique $n$-grams and $f_w$ are the corresponding frequencies constructed as described above. The parameter $n$ is called the \emph{thickness} of the dictionary. The set $\Omega = \{w\}$ is called the \emph{support} of the dictionary.
\end{definition}

It should be said that any text $\mathsf{T}$ could be unambiguously converted into a frequency dictionary; an inverse does not hold true, in general. Indeed, a set $S(l) \ni w$ of strings (of the given length $l$) assigned with the positive real numbers $f_w$ so that \[\sum_{w \in S} f_w = 1\] may correspond to a neither text. Since we have no aim to address a problem of a reconstruction of entire text from a dictionary, we shall not consider this issue, further.

A dictionary $D(n)$ of the thickness $n$ can be naturally projected into the frequency dictionary $D_1(n)$ of thickness $n-1$ bearing $(n-1)$-grams and their (reciprocal) frequencies. More generally, we can compute the dictionary $D_{k}(n)$ bearing $(n-k)$-grams with the reciprocal frequencies. It is a straightforward procedure that calculates all $(n-k)$-grams and their frequencies not from the original text $\mathsf{T}$ but rather from $D(n)$ with proper counting of the corresponding frequencies. This procedure uniquely defines the operator $\mathbb{P}_k{:}\ D(n) \to D_k(n)$ for $k$ ranging in $0,\dots, (n-1)$; here $\mathbb{P}_0{:}\ D(n) \equiv D(n)$. Further, such downward transformation will be denoted with a lower index; thus, $D_k(n) = \overline{D}(n-k)$ is indeed a frequency dictionary of the thickness $n-k$.

The inverse upward operator $\mathbb{L}_k{:}\ D(n) \to D^k(n)$ makes a frequency dictionary $D^k(n)=\overline{D}(n+k)$ of the thickness $n+k$ from the dictionary $D(n)$; here $k>0$ is an arbitrary positive integer. One can easily see that $D^k(n)$ is not uniquely defined; a family of different dictionaries $\{\overline{D}(n+k)\}$, instead. Any dictionary from the family yields the original frequency dictionary $D(n)$ due to an operator $\mathbb{P}_k$ execution: $\mathbb{P}_k[\overline{D}(n+k)] \to D(n)$, $\forall \overline{D}(n+k) \in \{\overline{D}(n+k)\}$. In such capacity, the operators $\mathbb{P}_k$ and $\mathbb{L}_k$ are not commutative ones: \[\left(\mathbb{P}_k\!\circ \mathbb{L}_k\right){:} \ D(n) \to D(n)\,;\quad \textrm{while}\quad \left(\mathbb{L}_k\!\circ \mathbb{P}_k\right){:} \ D(n) \to \ ?\]

To address this problem, one has to choose some peculiar frequency dictionary $\widetilde{D}(n+k)$ from the family $\{\overline{D}(n+k)\}$ of the extended ones. It should be kept in mind, that the family $\{\overline{D}(n+k)\}$ consists of various frequency dictionaries $\overline{D}(n+k)$, and the natural one $D(n+k)$ is among them. Here the maximum entropy principle may bring a solution, see \cite{n1,n2,n3,kitai} for the details and proofs. A brief outline of these results follows in the subsection \ref{rekonst}.

It turns out that the comparison of the same-thickness dictionaries $D(n)$ and $D^k(n-k)$ (i.\,e. extended dictionary vs. the natural one) provides the grounds for useful insights into statistical properties of the text $\mathsf{T}$, which are not readily accessible by other means.

Note that the original works \cite{n1,n2,n3,kitai} considered circularly looped input texts for the dictionaries generation. Here we can not require any periodicity of the input text $\mathsf{T}$ because it will create artificial connection between otherwise disconnected trading days at the beginning and at the end of the analyzed time interval. The absence of the loop will create a complication to be discussed later but for now we will just ignore it. The approximation by the results from the looped texts improves as $|\mathsf{T}| \to \infty$.

\subsection{Reconstructed Dictionary and the Information Valued $n$-grams}\label{rekonst}
Again consider an input text $\mathsf{T}$ defined over a finite alphabet $\aleph$. We can construct a sequence of dictionaries $D(j)$ of increasing thickness $j$:
\begin{multline}\label{zep}
D(1) \leftrightarrow D(2) \leftrightarrow \ldots \leftrightarrow D(j)\leftrightarrow\\ \leftrightarrow
D(j+1)\leftrightarrow \ldots \leftrightarrow D(L)\,.
\end{multline}
The projection operator $\mathbb{P}_k$ (arrows pointing left in \eqref{zep}), i.\,e., the transition $D(j) \mapsto D(j-1)$ is unambiguous. The opposite operator (that is $\mathbb{L}_k$) is ambiguous, generally, since an $n$-gram $w$ may have multiple valid continuations (not more than the cardinality of $|\aleph|$). 

A \emph{valid $1$-lift} is a transformation $L_1{:}\ D(j) \mapsto W(j+1)$ so that $W(j+1)$ is a dictionary of thickness $n+1$ and $P_1{:}\ W(j+1) \to D(j)$. So, by definition, a valid $1$-lift $L_1$ satisfies $P_1 \circ L_1 = I$, where $I$ is the identity mapping of $D(j)$. Thus, a lifted (extended) dictionary consists of $n$-grams $w \in \Omega$ extended by adding a prefix or a suffix of length $1$ in the way that the projection of that former yields the original frequency dictionary $D(n)$. Obviously, adding an infix to the original $n$-grams one may not get a valid lift. 

In other words, each combined set $f^{\ast}_{\nu_1\nu_2\nu_3\ldots \nu_{q-1}\nu_q\nu_{q+1}}$
of the extended $n$-grams must satisfy the constraint
\begin{multline}\label{cnst}
\sum_{\nu_{q+1}} f^{\ast}_{\nu_1\nu_2\nu_3\ldots \nu_{q-1}\nu_q\nu_{q+1}} =
\sum_{\nu_{q+1}} f^{\ast}_{\nu_{q+1}\nu_1\nu_2\nu_3\ldots \nu_{q-1}\nu_q} =\\ =
f_{\nu_1\nu_2\nu_3\ldots \nu_{q-1}\nu_q}\; ,
\end{multline}
where $f_{\nu_1\nu_2\nu_3\ldots \nu_{q-1}\nu_q}$ is the frequency of an $n$-gram $w \in D(q)$ in the original frequency dictionary $D(q)$. Linear constraints \eqref{cnst} eliminate some of the possible extensions for the original $n$-grams, but still do not define the lift uniquely.

As the final step to define the lift uniquely we shall use the maximum entropy principle:
\begin{equation}\label{me}
\max_{j} \left\{- \sum_{w^{\ast}} f^{(j)}_{w^{\ast}} \ln
f^{(j)}_{w^{\ast}} \right\}\,.
\end{equation}
Here $w^{\ast} = \nu_1\nu_2\nu_3\ldots \nu_{q-1}\nu_q\nu_{q+1}$ denotes an $n$-gram satisfying the linear constraint \eqref{cnst}, and $j$ enlists the versions of feasible extensions. The maximum-entropy dictionary $\widetilde{D}(q+1)$ satisfying both \eqref{cnst} and \eqref{me} exists always, since the set of the dictionaries to be constructed from the given one is finite.

The frequency of the $n$-grams in the max-entropy lift $\widetilde{w} \in \widetilde{D}(q+1)$ could be computed explicitly using La\,Grange multipliers method \cite{n1,n2,n3,kitai}. It is determined by the expression
\begin{multline}\label{vosst}
\widetilde{f}_{\nu_1\nu_2\nu_3\ldots \nu_{q-1}\nu_q\nu_{q+1}} 
=\\ =\frac{f_{\nu_1\nu_2\nu_3\ldots \nu_{q-1}\nu_q} f_{\nu_2\nu_3\ldots
\nu_{q-1}\nu_q\nu_{q+1}}}{f_{\nu_2\nu_3\ldots \nu_{q-1}\nu_q}}\;.
\end{multline}
Surprisingly, the expression \eqref{vosst} coincides with the Kirkwood's approximation \cite{kirk} (an absence of the ``interaction'' via the third ``particle'' makes it the exact solution of the problem here). Similarly, the maximum entropy principle~\eqref{me} allows to reconstruct the dictionary $\widetilde{D}(q+l)$ for $l$-lifts of any $l>1$, see \cite{n1,n2,n3} for details.

The $1$-lift to a thicker dictionary via \eqref{vosst} yields the dictionary that bears no ``additional'' information, external with respect to that one contained in the original dictionary. It consists of the $n$-grams of the length $q+1$ that are the most probable continuations of the strings of the length $q$. The lifted dictionary $\widetilde{D}(q+1) = \mathbb{L}_1{:}\ D(q)$ contains all the strings that occur in the original dictionary $D(q+1)$ and, possibly, some other ones. For any $q$, $q \geq 1$ the following inequality of the entropy: $$S\left[\widetilde{D}_{q+1}\right] \geq S\left[D_{q+1}\right]$$ holds true.

The maximum entropy approach could be easily generalized for valid $l$-lifts, $l>1$, again yielding a unique solution. Everywhere below we shall focus on $1$-lifts, only; besides, no other techniques of the lifting would be considered, but the max-entropy lift.

\subsection{Information Capacity of a Text}\label{info_alph}
Here we outline the idea of the information valuable $n$-grams (see sec.~\ref{rekonst}). Consider two sets of the frequency dictionaries of increasing length: the former consists of the entities constructed directly from the input text (the natural dictionaries) \[D(1) \leftrightarrow D(2) \leftrightarrow \ldots \leftrightarrow D(j)\leftrightarrow
D(j+1)\leftrightarrow \ldots \leftrightarrow D(L)\] and the latter \[\widetilde{D}(2) \leftrightarrow \widetilde{D}(3) \leftrightarrow \ldots \leftrightarrow \widetilde{D}(j)\leftrightarrow \widetilde{D}(j+1)\leftrightarrow \ldots \leftrightarrow \widetilde{D}(L)\] consists of the lifted dictionaries. Here we assume that $\widetilde{D}(j)$ is always lifted from $D(j-1)$, $j = 2,\dots, L$. 

\begin{definition}
\emph{Information capacity} $\overline{S}_j$ of a natural dictionary $D(j)$ is the mutual entropy 
\begin{equation}\label{infocap}
\overline{S}_j = \sum_{w\in \Omega} f_w \ln \left(\dfrac{f_w}{\widetilde{f}_w}\right)
\end{equation}
of the natural dictionary $D(j)$ calculated against its lifted-up entity $\widetilde{D}(j)$ derived from the dictionary $D(j-1)$. 
\end{definition}

This definition is applicable to any valid lifts. For the case of~\eqref{vosst} (max-entropy lift), the information capacity could be easily determined:
\begin{equation}\label{infocap:1}
\overline{S}_j = 2S_{j-1} - S_j - S_{j-2}\quad \textrm{and} \quad\overline{S}_2 = 2S_1 - S_2\,,
\end{equation}
where $S_j$ is the absolute entropy of the natural dictionary $D(j)$. A generalization of this expression for any $\widetilde{D}(q) = \mathbb{L}_k\circ \mathbb{P}_k{:}\ D(q)$ looks as \[\overline{S}_q = kS_{q-k} - S_q - S_{q-k-1}\quad \textrm{and}\quad \overline{S}_q = qS_1 - S_q\,.\]

\subsection{Information Valuable (Divergent) $n$-grams}\label{infvalstrings}
Consider again the information capacity~\eqref{infocap}. Sufficiently close values of natural frequencies $f_w$ and lifted frequencies $\widetilde{f}_w$ of the same $n$-gram $w$ make smaller contribution (per $n$-gram) to the overall value of the sum, while the $n$-grams with greater deviation provide greater-than-average contribution. This observation motivates the following:
\begin{definition}\label{ivw}
\emph{Information valuable $n$-gram} $\widehat{w}$ (an element of the frequency dictionary $D_j$) is an $n$-gram satisfying \[ |\log f_{\widehat{w}} - \log \widetilde{f}_{\widehat{w}} | > \log \alpha \,,\] where $1 \geq \alpha > 0$ is the information value threshold.

We will also call such $n$-grams \emph{$\alpha$-divergent} $n$-grams, or \emph{divergent} $n$-grams when parameter $\alpha$ is obvious from the context or its specific value is not important. 
\end{definition}

The complement to the subset of the divergent $n$-grams is the subset of \emph{$\alpha$-ordinary} $n$-grams (or just \emph{ordinary} $n$-grams). If $\alpha=1$ then all the $n$-grams within the dictionary $D_j$ are divergent with the exception of those whose lifted frequency exactly equal the natural one. Usually such $n$-grams occur only for sufficiently long input texts and sufficiently great $n$.

The choice of the parameter $\alpha$ in the definition~\ref{ivw} depends on a practical application. Making $\alpha$ large enough, so that the count $C_w$ of at least some of the divergent $n$-grams $w$ found in the dictionary is greater than 1, provides a reasonable guideline to setting minimal practical threshold. Such choice ensures that found divergent $n$-grams are not all degenerate (i.\,e. produced by combination of short input text and large $n$). It was said above such unique degenerate $n$-grams normally should be excluded from the analysis.

There are several issues stemming from the finiteness of the length of an input text. Since we shall use the divergent words as a tool in our studies of the financial time series, we should pay more attention on their features and obstacles arisen from the finite sampling, and discuss them immediately.

\subsubsection{Calculation of the Information Capacity: Looped vs. Finite Inputs}
The first issue we have to deal with is that some of the formulae derived for the looped or infinite input text will not hold true for finite texts absolutely, but approximately only. In particular, \eqref{infocap:1} would not be valid for information capacity. The origin of this formula shows that it relies on the exact equity of the sums in~\eqref{cnst} obtained due to the summation of frequencies of the $n+1$-grams, from both ends of a word. This assumption holds true for looped or infinite input text, only. Hence for practical calculations we should use the direct formula \eqref{infocap}.

\subsubsection{Noise Barrier}
Another point is related to the noise resulted from the finiteness of a length of an input text to affect the figures of information capacity. We have already touched this subject briefly earlier but a consideration of some more details would be helpful for further analysis of financial data (see Section~\ref{csaracf}). 

To simplify the issue, consider a binary alphabet with proximal probabilities of both symbols. The total number of \emph{different} $n$-grams of length $n$ is then $2^n$. There are total $L-n+1$ of \emph{all} $n$-grams of the length $n$ in the input text of the length $L$. For $L \gg n$ we can take the number of $n$-grams $\approx L$. If $L=2^n$ then each occurrence of each $n$-gram is ``critical'' in a sense that every difference of lifted dictionary from the natural one will ``amplify'' the random nature of the input text.

When dictionary thickness exceeds $\log_2 L$, some of the $n$-grams will not be present at all (go ``extinct''). This follows in a degeneration of the information capacity as many of the longer $n$-grams will be lifted from the shorter ones uniquely and other will not be present. The number of uniquely reconstructed $n$-grams grows up as the ratio $L/n$ gets smaller. This results in the bell curve of information capacity~\eqref{infocap}. The peak is located at the value of $n$ close to $j_{\max} \approx \log_2 L$. 

Similarly for the alphabets of cardinality $\|\aleph\| = k$ the figure $j_{\max} \approx \log_k L$ approximates the location of the peak. Again, this approximation holds better when the probability density of the letters in the alphabet gets closer to the uniform one, hence making the value $j_{\max}$ smaller when the distribution is far from the uniform. We will call the figure of $j_{\max}$ \emph{the noise barrier} for the given input text length.

\subsubsection{Normalized Information Capacity}
To separate the virtual signal in the information capacity $\overline{S}_n$ of the input text $\mathsf{T}$ from the noise, we need to compare it to the expectation $\mathsf{E}(S'_n)$ of the information capacity $S'_n$ calculated from the randomly generated surrogate texts $\mathsf{T}'$. Random texts $\mathsf{T}'$ must have the same length and yield the same probabilities of the alphabet letters as $\mathsf{T}$ does. The figure of an absolute difference of the values $S_n$ and $S'_n$ are not as useful, as the normalized one by the standard deviation $\sigma(S'_n)$ of the information capacity of the corresponding random input texts $\mathsf{T}'$. This normalization gives $\sigma$-distance from the purely random signal. 

\begin{definition}\label{norm_inf_cap}
The \emph{normalized information capacity} of the input text $\mathsf{T}$ is defined as
\begin{equation}\label{norm_infocap}
S^{\ast}_n = \frac{\overline{S}_n - \mathsf{E}(S'_n)}{\sigma(S'_n)}\,,
\end{equation}
where $\overline{S}_n$ is the information capacity of the original input text; $\mathsf{E}(S'_n)$ and $\sigma(S'_n)$ are the expectation and the standard deviation correspondingly of the information capacity of the random input text $\mathsf{T}'$ so that $\mathsf{E}(D(1)(\mathsf{T}'))=D(1)(\mathsf{T})$. The last condition means that the the letters in the source of the random texts $\mathsf{T}'$ are distributed in the same way as for the original text $\mathsf{T}$.
\end{definition}

In practice, to estimate the values of $\mathsf{E}(S'_n)$ and $\sigma(S'_n)$ one should go through the following steps using Monte-Carlo method:
\begin{list}{\arabic{N}\,)}{\topsep=0cm \itemsep=0cm \listparindent=1cm \parsep=0pt \usecounter{N}}
\item Compute probability distribution $D(1)$ of the letters in the input text;
\item Generate a set $\{\mathsf{T}'_k\}$ $k=1, \dots , M$ of sufficiently large number $M$ of the random texts $\mathsf{T}'$ of the same length generated using probabilities $D(1)$;
\item For $\mathsf{T}'$ estimate $\mathsf{E}(S'_n)$ and $\sigma(S'_n)$.
\item Calculate the normalized value~\eqref{norm_infocap}.
\end{list}
The complexity of this method is obviously exponential, what makes the parameter $|\mathsf{T}|=N$ that is the length of an input text important with respect to the computational costs.

\section{Case Studies, Aliasing and the Relation to ACF}\label{csaracf}
We start the analysis with a broad index Russel 2000 that may represent to certain extent a typical behaviour of broad market returns. Then we move to the behaviour of historical returns of Bank of America (ticker BAC) while touching on some other tickers. In further discussion all the input texts have been generated via binary quantization. The typical time window for the analysis was three years. 

\subsection{Historical Returns of Russel 2000 (\^{ }RUT); ACF vs. Information Capacity}
\begin{figure}[!b]
\centering
\includegraphics[width=9.1cm]{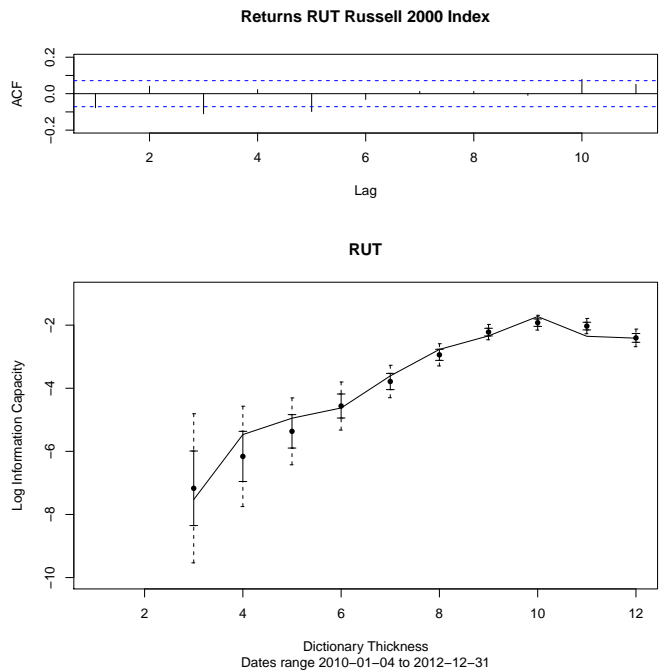}
\caption{\footnotesize Upper chart: auto-correlation (ACF) of daily $\log$-returns. Lower chart: solid line shows $log$-information capacity, dots show $\mathsf{E}(S'_n)$ (the expectation of the information capacity), solid vertical bars mark $\sigma(S'_n)$ and dotted vertical bars mark $2 \sigma(S'_n)$ as in Definition \ref{norm_inf_cap}.}
\label{figure:example}
\end{figure}

Figure~\ref{figure:example} shows $\log$ of the absolute values of the information capacity calculated for the ticker \^{ }RUT. We used logarithmic scale for the plot since the information capacity has different order of magnitudes at the different dictionary thickness figures. There were approximately 750~trading dates worth of data in the represented time window. It gives the noise limit value between 9 and~10. The peak of information capacity is observed close to the dictionary thickness~10, as expected. 

There are few more important observations coming from Fig.~\ref{figure:example}. The first observation indicates that the behaviour of the wide market returns is quite close to the noise and does not exhibit any dramatic deviations as the solid line lies within single $\sigma$ from the noise signal.

The second observation is that auto-correlation (ACF) and the information capacity plotted together show no obvious connection. This suggests that $n$-grams-based analysis is not directly tied with ACF-based one and represents a new statistical aspect of the input text. Yet there must be some connection between information capacity and ACF. The following simple argument supports this hypothesis. Consider periodic input text $T$ with the period $L$ and the length $|T| \gg L$. Obviously, the information capacity of such periodic text degenerates (becomes 0) on the dictionary thickness $ > L$. And the ACF has maximum at the value $L$. The authors will revisit the connection between information capacity and auto-correlation in the future works. 

\begin{figure}[!h]
\centering
\includegraphics[width=9.1cm]{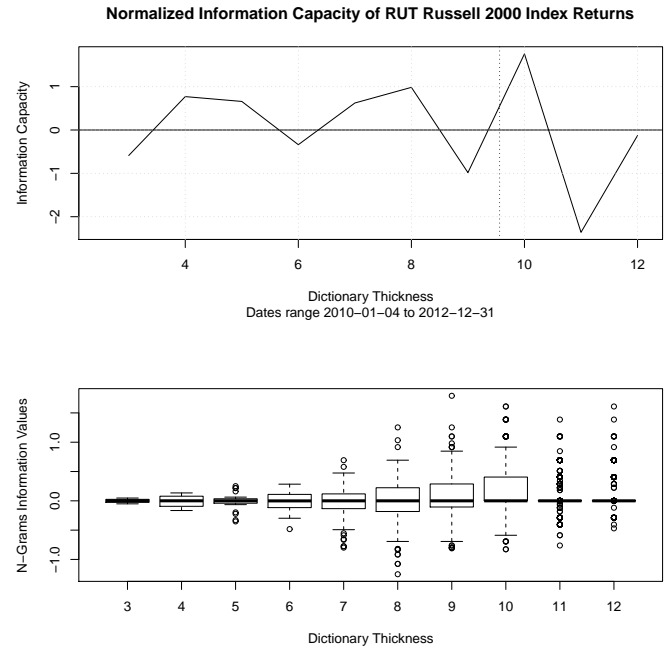}
\caption{\footnotesize Upper chart: normalized information capacity. Lower chart: box plot for the $n$-grams distribution of the input text. The vertical dotted line indicates the noise limit of the input text.}
\label{figure:example2}
\end{figure}

A quick comparison of $\log$-information capacity (Fig.~\ref{figure:example}) and normalized information capacity provided in Fig.~\ref{figure:example2} shows the advantages of the normalized representation for the information capacity. Additionally, Fig.~\ref{figure:example2} shows box-plot for the $n$-grams in the different thickness dictionaries. It is interesting to confirm visually that the $n$-grams distribution degenerates into multiple outliers and the condensed central peak for the values above noise limit (vertical dotted line in Fig.~\ref{figure:example2}). 
The overall explorative analysis using information capacity for the ticker \^{ }RUT indicates quite good correspondence with the random market hypothesis.

\subsection{Case studies: Starbucks Corp. (ticker SBUX) and Bank of America Corp. (ticker BAC)}
Starbucks Corp. shows more interesting pattern, which is less in line with the hypothesis of random behaviour of returns. This manifests in high deviation from the information capacity of the corresponding noise and somewhat slower degeneration of the $n$-grams distribution for the values $n$ close to the noise limit. This last observation is supported by the box-plots. Another example is provided by Bank of America Corp. (ticker BAC). Its information capacity exhibits significant deviation from the average information capacity for $n = 7$.

The less noisy behaviour of the tickers for particular companies was discovered in many more cases. This hardly is unexpected. The behaviour of the returns for particular companies is heavily influenced by the events significant for the company (e.\,g. earning reports, other announcements from the company and the competition, etc). Thus, particular company returns are less prone to the averaging to the pure noise, unlike the wide market returns where such events are more spread out in time and diluted by the large number of the index components.
\begin{figure}[t]
\centering
\includegraphics[width=9.1cm]{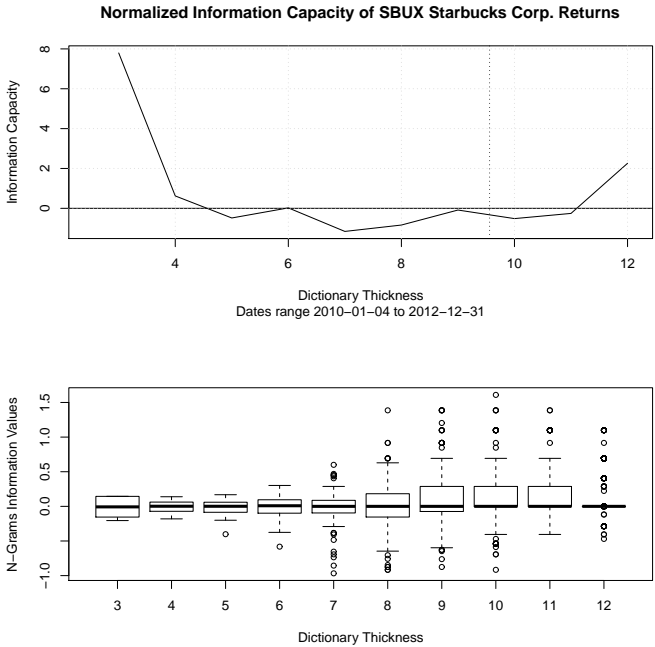}
\caption{\footnotesize Starbucks Corp. normalized information capacity shows significant deviation from that one of the corresponding random noise on the dictionary thickness $n=3$. See also discussion of the aliasing noise for small values of $n$ in subsection \ref{aliasing}.}
\label{figure:example3}
\end{figure}

\begin{figure}[!b]
\centering
\includegraphics[width=9.1cm]{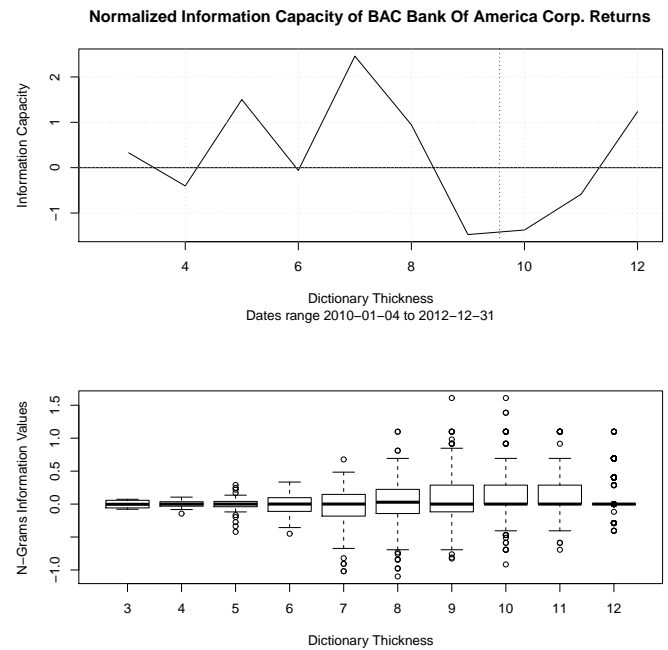}
\caption{\footnotesize Ticker BAC (Bank of America Corp.): the information capacity is outside of $2\sigma$ distance from the corresponding noise for $n=7$. }
\label{figure:example3a}
\end{figure}

\subsection{Aliasing}\label{aliasing}
Even though the signal for SBUX shown in Fig.~\ref{figure:example3} at $n=3$ is unusually strong, we hardly can attribute it entirely to some source of meaningful information. The reason for that comes from the very limited cardinality of the dictionary $D_3$. For the binary alphabet $|D_3|=2^3=8$, i.\,e. we have only 8 different $n$-grams available for the analysis of information capacity. Hence, all of the sequences of the length~3 contained in the original input series of the returns are mapped via quantization mapping to only~8 available $n$-grams. Thus truly significant triplets of the market days for the ticker get binned together with the ordinary ones only due to the lack of the unique $n$-grams in the target quantization space. This effect is similar to the aliasing in image and signal processing when the limited target quantization space generates unwanted artifacts and leads to the additional information loss. 
\begin{figure*}
\centering
\includegraphics[width=18.6cm]{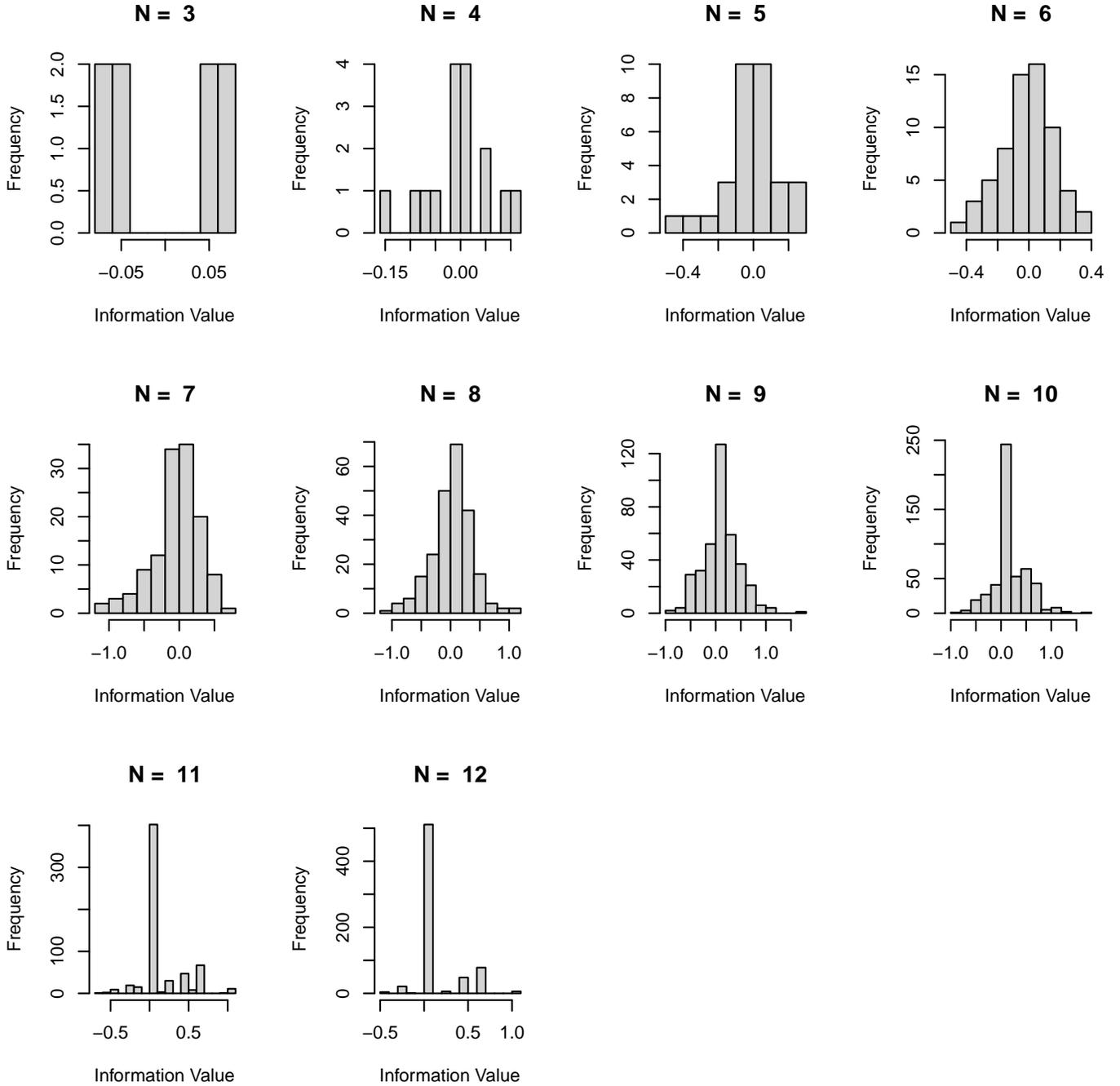}
\caption{\footnotesize Histograms of $n$-grams distribution for Bank of America Corp. (BAC) for different values of $n$.}
\label{figure:example5}
\end{figure*}

Such aliasing also makes the choice of the $\alpha$-divergent $n$-grams for smaller values of $n$ difficult or even impossible because the comprehensive identification of the threshold value $\alpha$ (see definition~\ref{ivw}) should result in non-unique $n$-grams occurrence. The example of less extreme behaviour of BAC illustrates strong aliasing at $n=3$ (see Fig.~\ref{figure:example5}). For $n=3$, the information capacity of $n$-grams is batched into two clusters on the extremes of the range, thus leaving very few choices for the value of $\alpha$ in $\alpha$-divergence. 

Finally, we excluded $n=2$ from our analysis completely because of the aliasing.

\subsection{Divergent $n$-grams: Distribution over Time}
\begin{figure*}
\centerline{\includegraphics[width=17.6cm]{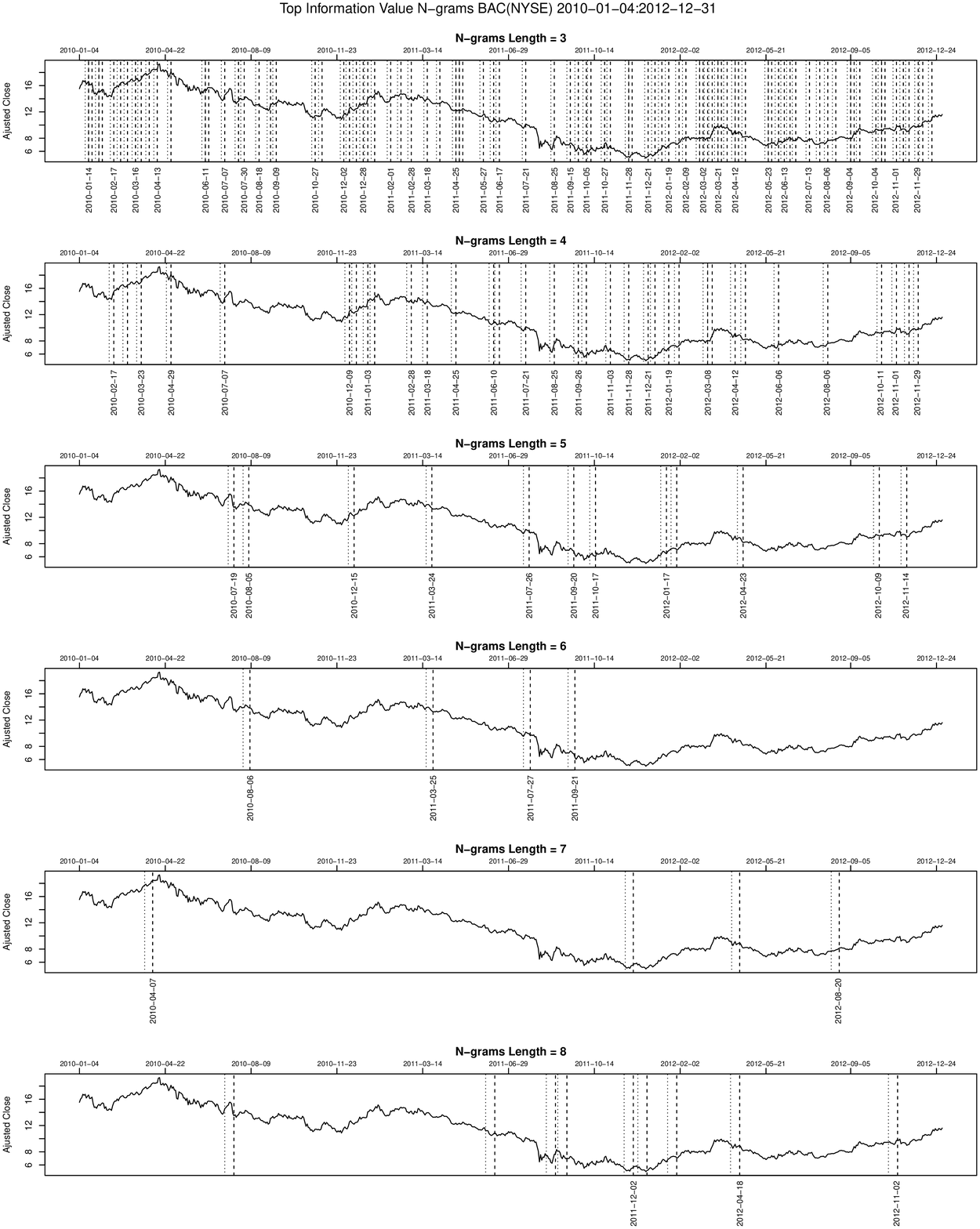}}
\caption{\label{figure:example4} Bank of America Corp.: distribution of divergent $n$-grams over time.}
\end{figure*}

The location of divergent $n$-grams on the time axis can be studied from the point of view of their possible connection to the market events or particular price action of the underlying ticker. Fig.~\ref{figure:example4} shows an overlay of such divergent $n$-grams for different value of $n$ over the price plot, for the ticker BAC. The divergent $n$-grams were picked in such a way that none of them was unique and they represent no less than~5\,\% of the total number of the $n$-grams. For smaller $n$ that resulted in higher percentage due to the aliasing discussed earlier in sub-section~\ref{aliasing}. 

Fig.~\ref{figure:example4} inspires several observations. The first one is that the location of the longer divergent $n$-grams on the time axis does not necessarily coincide with the location of the shorter ones. In other words, there is no incidence of the divergent $n$-grams of various length. It may come from the instability of divergent $n$-grams against the noise of different origins. To mitigate this effect, we may consider studying only stable divergent $n$-grams i.\,e. $n$-grams that possess filtration property. It is still unclear if such stable divergent $n$-grams can be used as technical indicators marking anything of interest like markets top, beginning or end of a trend, etc.

Another possible reason for the lack of filtration of divergent $n$-grams can be expressed as a hypothesis regarding the role of divergent $n$-grams in the indirect information exchange between market participants. Filtration here means an identification of the incident divergent $n$-grams of increasing length $n$. The shorter ones are responsible for the shorter time range and are generated by interaction of short-horizon market participants. The longer ones are possibly related to the activity of longer horizon investors. The absence of the information exchange on the shorter periods of time suggests that such exchange is also unlikely on longer time periods that contain the shorter one in question. Such ordinary periods of time devoid of divergent $n$-grams would represent ``business as usual''. Thus we may expect ``filtration'' of such ordinary stretches of time from smaller $n$ to larger $n$. This hypothesis is in reasonable correspondence with the plots in Fig.~\ref{figure:example4}.

\subsection{Divergent $n$-grams and Market Events}
\begin{figure*}
\includegraphics[width=18.5cm]{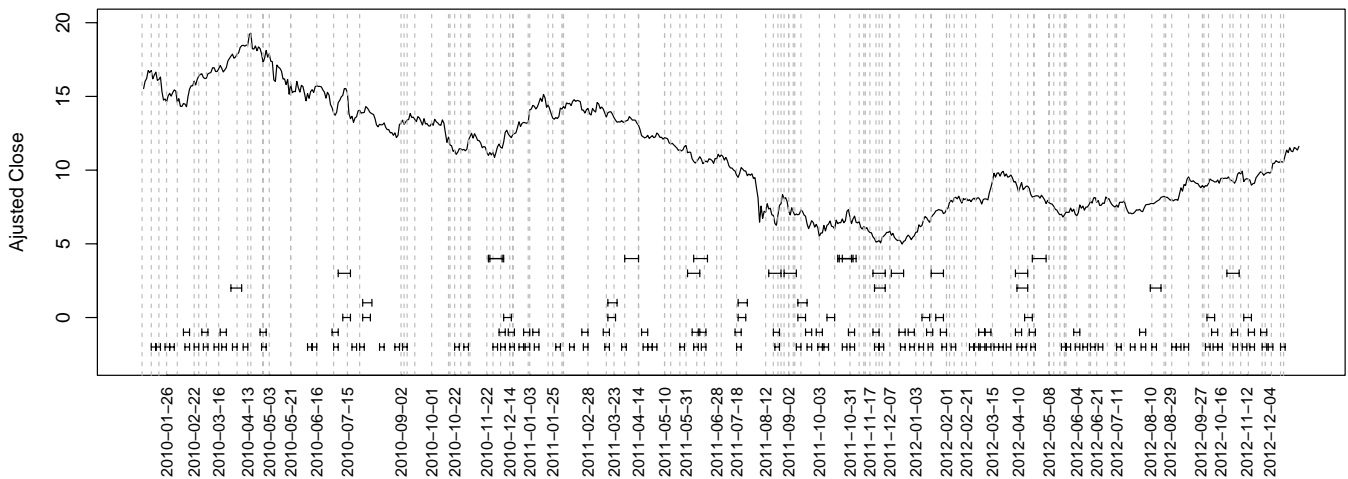}
\caption{\label{figure:exampleNews} Events and announcements of Bank of America Corp. together with the divergent $n$-grams. Horizontal bars represent non-unique divergent $n$-grams (5\%-percentile). Vertical dotted lines are the official company events and announcements.}
\end{figure*}

A hypothesis that the divergent $n$-grams are connected to particular market events can be tested directly. Fig.~\ref{figure:exampleNews} shows an overlay of the official company events and announcements over price chart and divergent $n$-grams for ticker BAC. The events were pulled off the official investors section of the website of Bank of America Corp. The absence of direct connection between events and divergent $n$-grams appears to be inconclusive. The official events are polluted with a stuff that are not really significant and may not include rumors, information leaks or external events significant for the company. The insider trading data published by the company may be also included into the analysis. It may be worth to attempt to filter out less significant official events and include important external ones in the further analysis. 
\begin{table}[ht]
\caption{Non-unique divergent $n$-grams (5\,\%-procentile) for two overlapping time windows. \textbf{I} stands for the period 2010-01$\div$2012-12; \textbf{II} stands for the period 2010-03$\div$2013-03} 
\centering 
\begin{tabular}{|c|c|c||c|c|c|}\hline 
$n$-grams & \textbf{I} & \textbf{II} & $n$-grams & \textbf{I} & \textbf{II} \\ [0.5ex]
\hline
3 & 011 & & 8 &00110111 & 00110111\\
  & & 111 && 11111000& 11111000\\
\cline{1-3}
4 & 0011 & 0011 & {} & 10000111 & 10000111\\
&&& & 10001001 & {}\\
\hline
5 & 11000 & 11000 &9 &111001000 & 111001000\\\cline{1-3}
6 & 110000 & 110000 &&101011100& 101011100\\\cline{1-3}
7 & 1011111 & 1011111 && 010111001& 010111001\\
{}& 0110111 & {} && 011100110& {}\\
{}&{}& 0100011 & & {}& 110010111\\\hline
\end{tabular}
\label{table:table1}
\end{table}

Alternatively, one can judge the significance of an event by the information value of the $n$-gram that it generates (or is preceded by). Both approaches require wider study on more diverse input data in order to rectify their methodology.

\subsection{Divergent $n$-grams: Stability with Respect to Time Window Shift}
The divergent $n$-grams computed for one time window do not have to coincide with those computed for different but overlapping time window. This can be called time-shift (in-)stability. It turned out that divergent $n$-grams are usually reasonably stable. Table~\ref{table:table1} illustrates this observation with the data for the ticker BAC. The binary alphabet is $\{0,1\}$ and represents up- and down- ticks. 

\section{Discussion}\label{discussion}
The paper presents a novel approach of a study of the dynamics of financial series through the technique of frequency dictionary analysis. To do that, one has to quantize the original market data into a symbol series. Neither the quantization, nor a frequency dictionary approach themselves make a novelty; a combination of these two ideas makes that latter. It should be said that a continuous analog of the approach present above is also possible, due to wavelet technique. Meanwhile, the continuous case is more complex and brings less understanding of the issue standing behind the observed data. Nonetheless, the choice of an alphabet is of key point here.

\subsection{Notes on the Choice of the Alphabet}
An exclusion of $0$ from the alphabet indexes may look clumsy, but it is so since we do not take into account ``neutral'' days on the market. However this follows the popular financial modeling methodology when only up and down ticks in prices are taken into account, but `zero' ticks are neglected. Evidently the described text generation is just an encoding of the net number of up or down ticks accumulated at the end of the day and is represented by Adjusted Close price.

The inclusion of $0$ in the alphabet may be justified when we are dealing with missing data, or temporary suspension from trading, or absence of trades on certain days for low liquidity assets, or the days when the market is closed. In all these cases $0$ will have special meaning that will complicate analysis. Yet it may be very useful as, e.\,g. Fridays and Mondays tend to show more significant market movements and the no-market days are not represented in $z(t)$ in any direct way. We leave these considerations outside of the current paper and focus on the simpler cases to build up our tools. 

\subsubsection{The Binary Alphabet and an ``Alphabet Staircase''}
The binary alphabet represents an interesting special case and is worth a more detailed discussion. Our primary subject of interest will be frequency dictionary $D(n)$ of $n$-grams naturally constructed from the original text. Obviously, the cardinality of the dictionary for sufficiently noisy texts grows exponentially with the linear growth of its thickness (which by the definition is equal to the length of the corresponding $n$-grams). Thus, we may expect that some ``essential'' features of the input text manifesting themselves through a frequency dictionary may be better studied on the logarithmic scale (the features grow exponentially while the input length grows linearly). To introduce this logarithmic we construct an ``alphabetic staircase''. It starts with the binary alphabet. The next step is ``quaternary'' alphabet, then ``octonary'' alphabet, etc. with thickness going over the powers of two: $2^k$, $k = 1, 2, \ldots$ forming an uprising ``staircase''.

The staircase construction leads to the termination problem: what $k$ should we use to stop the process? It is easy to see that quite soon, depending on the length of the input texts, the alphabet becomes redundant. This results in the degeneracy of a dictionary with most of the $n$-grams encountered only once. Unfortunately, until the degeneration becomes complete, there is no immediately obvious way to stop at any specific dictionary thickness. However we can consider some heuristic techniques. We present one of them further (see subsec.~\ref{info_alph}), and another one may work as follows.

The growth of $n$-grams length extracted from a fixed-length input text also results in an increased number of the elements of a dictionary encountered only once. Such unique $n$-grams do not contribute to the meaningful analysis. So, we can exclude unique $n$-grams and focus on those with multiple occurrences in the corresponding dictionary $D(n)$. Such non-unique $n$-grams form a sub-dictionary $\widetilde{D}(n) \subseteq D(n)$. 

Our next step in evaluation of the optimal termination of the the dictionary thicknesses staircase is based on the the entropy estimation of the dictionaries $\widetilde{D}(n)$, within the geometrically progressing series of dictionaries of thickness $\widetilde{D}(2^k)$. Namely, we expect that entropy of $\widetilde{D}(n)$ will reach maximum at some $n=2^k$: \[\max_{k} \left\{ -\hspace{-9pt}\sum_{w \in \widetilde{D}(2^k)} \hspace{-7pt}f_w \ln f_w \right\} \,.\]
Here $f_w$ is the frequency of an $n$-gram $w$ in the dictionary $D(n)$. The value of $arg_{\max}$ of the expression above will represent the optimal stopping $k$ for the dictionaries staircase.

\subsection{A Note on the Choice of Quantization Mapping}\label{ideas}
As we have already mentioned, there is no obvious choice of the quantization mapping $Q{:}\ R \to A$. However there are some natural considerations that can be taken into account. Obviously, smaller alphabets (coarser quantization) produce simpler texts but that may result in the loss of useful information contained in the original series $p(t)$. On the other hand, an over-abundant alphabet will result in mostly unique $n$-grams, even for small $n$. Hence the alphabet size and the input series length must be balanced.

It is also worth mentioning that for the quantization and further analysis we replace the absolute price series $z(t)$ with the (log-)return time series $p(t)$. We based such a choice solely on simplicity. Instead, one may consider excessive returns (i.\,e. returns on a ticker compared to the market returns), or deviation from moving average $z(t)-\mathsf{MA}(m, z(t))$, or pretty much any other ``reasonable'' function. Taking into account that there are hundreds of already adopted so called technical indicators, which values can be represented by a single number on a day $t$, the choice can be overwhelming. We will revisit this issue in the future works but will focus here on the simple method based on returns.

A somewhat more formal treatment of the choice of quantization may include the following problem: Given a class of quantization mappings, find an ``optimal'' representative in that class.

The definition of optimality can be constructed using equations~\eqref{infocap:1} and general observations regarding relative entropy. It is easy to see that $\overline{S}_j$ is always bell-shaped, as a function of $j$. This fact results from a degeneracy of a frequency dictionary $D_j$, as $j \rightarrow \infty$: the $n$-grams occurring only once become dominant. For any input text $\mathsf{T}$ and any quantization $\mathfrak{Q}$, there always exists such $j^{\ast}$ that $\overline{S}_{j^{\ast}}$ is maximal. Maximizing $\overline{S}_{j^{\ast}}$ over the elements in the selected class of the quantizations will result in an optimal quantization: 

\begin{definition}
\emph{Information-optimal quantification} is the quantification $\mathfrak{Q}^{\ast}$ that yields the maximal value of \[j^{\ast}_{\max} = \max_{\{\mathfrak{Q}\}}\{j^{\ast}\}\,.\]
\end{definition}

\section{Conclusion}
The proposed approach to the analysis of time series based on the information capacity of $n$-grams extracted from the corresponding discrete texts can provide potentially valuable new tools and statistical metrics. The paper discussed several possible applications of the new approach and illustrated them with case studies of the actual market data. Also we discussed limitations of the approach and quantified them by estimating the noise limit and the aliasing effect. The connection of divergent $n$-grams with market events or their value as technical indicators remains a topic open for deeper investigation. However, even superficial observations with limited data and binary quantization hint at the connection of the divergent $n$-grams with implicit information exchange between market participants. Such hypothesis, if proven to be right, can have significant value for the market analysis. 

We also outlined an interesting theoretical problem of estimating expectation and variance of the information capacity of the noise. Analytic solution to such problem will be useful in many areas where $n$-grams analysis plays significant role.

\end{document}